\documentclass[11pt]{article}
\usepackage{amsmath,graphicx}
\usepackage{csvsimple}
\usepackage{booktabs,tabularx}
\usepackage{siunitx}
\usepackage{INTERSPEECH_mod}

\title{CARNATIC RAGA IDENTIFICATION SYSTEM USING RIGOROUS TIME-DELAY NEURAL NETWORK}
\name{Sanjay Natesan$^1$, Homayoon Beigi$^{1,2}$}
\address{$^1$ Department of Computer Science, Columbia University, New York, USA 10027 \\
  $^2$ Recognition Technologies, Inc., South Salem, USA 10590}
\email{$^1$sanjay.natesan@columbia.edu, $^2$beigi@recotechnologies.com}

\begin{document}
%
\maketitle
\begin{abstract}
Large scale machine learning-based Raga identification continues to be a nontrivial issue in the computational aspects behind Carnatic music. Each raga consists of many unique and intrinsic melodic patterns that can be used to easily identify them from others. These ragas can also then be used to cluster songs within the same raga, as well as identify songs in other closely-related ragas. In this case, the input sound is analyzed using a combination of steps including using a Discrete Fourier Transformation and using Triangular Filtering to create custom bins of possible notes, extracting features from the presence of particular notes -- or lack thereof.

Using a combination of Neural Networks including 1D Convolutional Neural Networks (conventionally known as Time-Delay Neural Networks) and Long Short-term Memory (LSTM), which are a form of Recurrent Neural Networks, the backbone of the classification strategy to build the model can be created. In addition, to help with variations in shruti, a long-time attention-based mechanism will be implemented to determine the relative changes in frequency rather than the absolute differences. This will provide a much more meaningful data point when training audio clips in different shrutis. To evaluate the accuracy of the classifier, a dataset of 676 recordings is used. The songs are distributed across the list of ragas. The goal of this program is to be able to effectively and efficiently label a much wider range of audio clips in more shrutis, ragas, and with more background noise.
\end{abstract}
\section{Introduction}
\label{sec:intro}
Carnatic music is a classical music tradition originating from South India, with roots dating back over 2,000 years. It is characterized by intricate melodies, rhythmic patterns, and improvisation, often accompanied by instruments like the veena, flute, violin, mridangam, and tanpura. Carnatic music is usually classified into three subparts: \textit{shruti}, \textit{laya}, and \textit{raga}. Shruti refers to the scale of the piece, and in the modern era, the 12 shrutis used per octave are congruent to the 12 notes used in the Western scales \cite{srimani}. However, traditionally an octave can instead be split into 22 shrutis, allowing for finer gradation of the pitch \cite{srimani}.

Laya refers to the repetitive rhythmic timing of a piece. In Carnatic music, laya is established using talam, which is the cycle of equidistant beats for a given song. One measure of talam, referred to as an avarthanam, represents the division of beats into clear patterns. The use of talam helps to indicate the pace of the song \cite{sridhar_swara}.

The final pillar of Carnatic music is the raga, a more complex version of the western melody. These ragas are built on a framework of seven primary notes (S, R, G, M, P, D, N), which correspond roughly to the Western Do, Re, Mi, Fa, Sol, La, Ti. These seven syllables, called swaras, can be split further into 16 notes separated by one half step, displayed in Table \ref{table:1} with their frequencies in C shruti. The table also shows the ratio of each note to the standard S note. This ratio stays constant regardless of the scale of the song. Note that in Carnatic music, certain pitches will overlap with each other. For example, R$_2$ $=$ G$_1$, R$_3$ $=$ G$_2$, D$_2$ $=$ N$_1$, and D$_3$ $=$ N$_2$. 

\begin{table}[h!]
\centering
\begin{tabular}{|c|c|c|}
\hline
Note & Frequency ($Hz$) & Ratio to S \\
\hline
S & $130.81$ & $1.0000$\\
\hline
R$_1$ & $138.59$ & $1.0595$ \\
\hline
R$_2$, G$_1$ & $146.83$ & $1.1225$ \\
\hline
R$_3$, G$_2$ & $155.56$ & $1.1892$ \\
\hline
G$_3$ & $164.81$ & $1.2599$ \\
\hline
M$_1$ & $174.61$ & $1.3348$ \\
\hline
M$_2$ & $185.00$ & $1.4143$ \\
\hline
P & $196.00$ & $1.4984$ \\
\hline
D$_1$ & $207.65$ & $1.587$ \\
\hline
D$_2$, N$_1$ & $220.00$ & $1.6818$ \\
\hline
D$_3$, N$_2$ & $233.08$ & $1.7818$ \\
\hline
N$_3$ & $246.94$ & $1.8878$ \\
\hline
\end{tabular}
\caption{Frequency of Notes in C Shruti and ratio between notes and S \cite{sridhar_raga}}
\label{table:1}
\end{table}

For each raga in Carnatic music, there is a corresponding ascending and descending melodic scale, refered to as an \textit{arohanam} and \textit{avarohanam}, respectively \cite{krishna}. Within each raga’s arohanam and avarohanam, the frequency of each swara must be greater than any swara that precedes it in the natural order described above (e.g. there can be no raga that has a combination of G$_1$ and R$_3$, or D$_2$ and N$_1$). However, there can be ragas with an arohanam or avarohanam that do not follow the natural swara order, but still follow the frequency order.

There are 72 ragas in Carnatic that are given the special classification of a Melakarta, or parent raga. Melakartas are defined as ragas that contain all seven swaras in the arohanam and avarohanam, with each swara sung at the same frequency in both the arohanam and avarohanam. The number seventy-two is thus obtained from all possible combinations of the variable swaras (R G M D N). Derived from these foundational melakartas are the Janya ragas (child ragas), which are characterized by missing or extra swaras in the arohanam and/or avarohanam 
(e.g. S R$_2$ G$_3$ P D$_2$ $\Bar{\text{S}}$) or the use of twisted progressions (e.g. S G$_3$ R$_2$ M$_1$ P N$_3$ $\Bar{\text{S}}$) \cite{sriram}. Janya ragas can consist of as few as 4 notes (not including the ending $\Bar{\text{S}}$ in the next octave to close the scale), and are estimated to number in the tens of thousands.

Carnatic music also includes a technique that separates it from many other forms of music, known as \textit{gamaka}. Gamaka is the modulation of a swara sung to provide additional beauty and variation. Seven types of gamaka are commonly used:
\begin{itemize}
    \item Jantai - the use of additional emphasis on a repeated note \cite{krishna}
    \item Jaru - the sliding from a slightly higher or lower frequency through the true frequency of the note \cite{krishna}
    \item Kampita - the rapid oscillation on a single note by touching the swara directly above and below it \cite{krishna}
    \item Khandippu - the descent from an initial swara to the final nonadjacent swara with brief intonations on swaras in between \cite{krishna}
    \item Odukkal - similar to jaru, shifting to the next swara in the raga before returning to the original swara \cite{krishna}
    \item Orikai - shifting to the next swara in the raga before descending to the swara below the original \cite{krishna}
    \item Sphuritam - starting a swara a higher frequency than natural and rapidly descending \cite{krishna}
\end{itemize}
 
Gamakas add an intricate level of detail to Carnatic music. However, they also provide an additional challenge to machine algorithms that traditionally rely on frequency identification. 

In this project, we will take a novel approach to raga recognition, using LSTMs and CNNs. Rather than trying to identify pre-determined phrases, we use a combination of neural networks with several layers to allow the machine to determine what patterns and phrases are relevant on its own. Further, this project will greatly expand the dataset to include other ragas and songs in a greater variety of samples, and will explicitly use relative frequency changes rather than absolute frequency changes to attempt to increase overall accuracy. The goal is to stay consistent with existing accuracies even as we increase the potential for variance by increasing the list of ragas to predict. The accuracy will be measured using the Loss Function, as has been done in previous research, to ensure direct comparisons can be made \cite{gulati_time}.

\section{State of the Art}
\label{sec:state}

Most existing research in machine learning methods for raga recognition is based on either Pitch Class (PC) identification, or frequency sequence. Some of the best performing PC id methods across both Carnatic and Hindustani (North Indian Classical) music \cite{chordia} have reached accuracy rates of $91\%$. However, PC id methods have been found to lose the intricacies of Carnatic music through their use of first order functions, which can often smooth over gamakas \cite{antiPC}. This will cause issues when more complex ragas are incorporated into the test set. In more current research, there have been attempts to overcome these deficiencies, such as through Hidden Markov Models \cite{hmm}, or using the arohanam and avarohanam as support metrics \cite{sridhar_raga}, among others.

In this project, we will be using the arohanam and avarohanam -- which, while critical, do not contain much expandable information -- in conjunction with actual audio clips to maximize the amount of possible reference points. We will also reduce the reliance on pre-processing that has become common in Carnatic raga identification research, to avoid the possible loss of complex features.

\section{Method}
\label{sec:method}

\begin{figure}[t]
    \centering
    \includegraphics[width=0.5\textwidth]{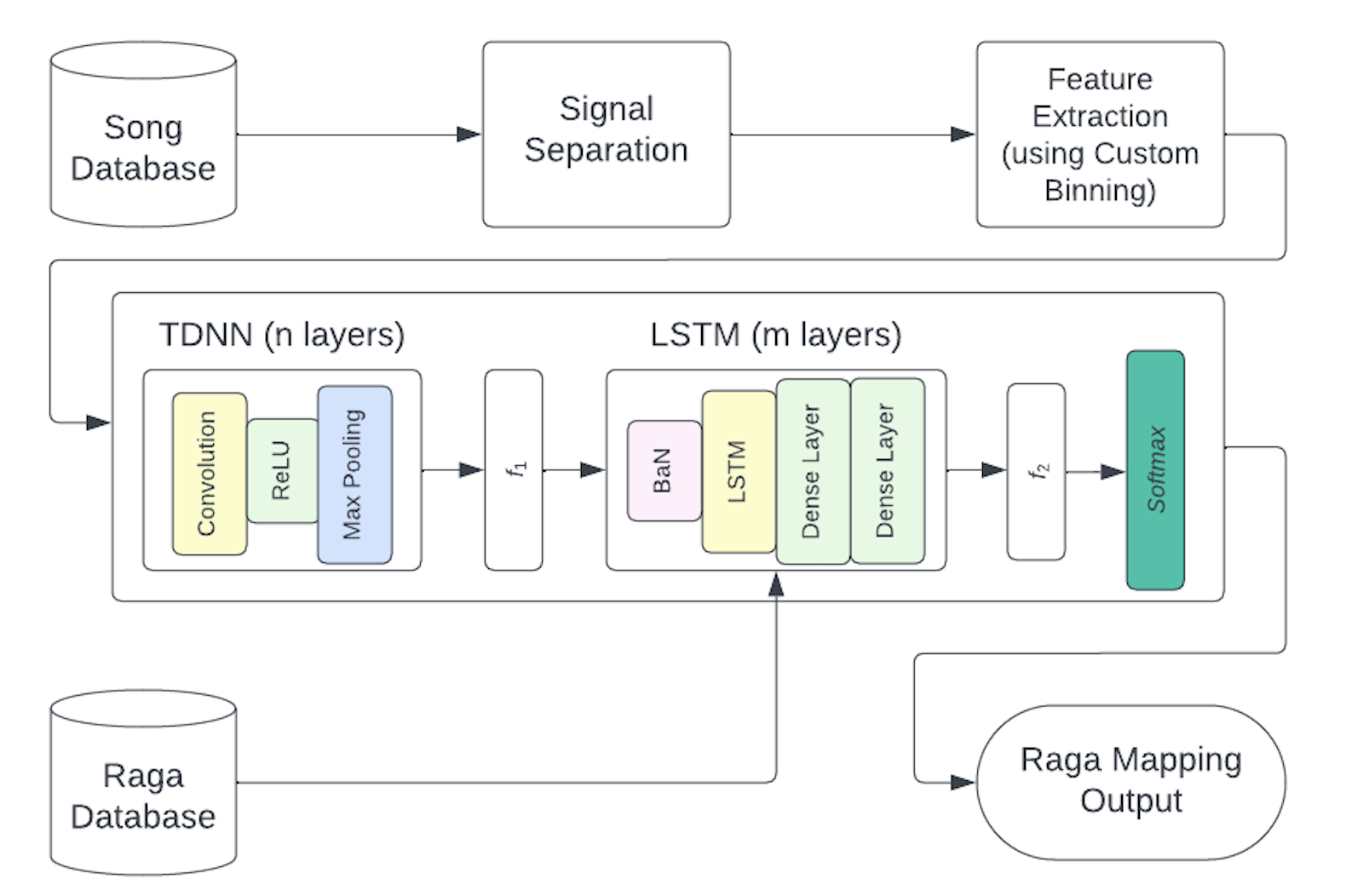}
    \caption{System Architecture}
    \label{fig:arch}
\end{figure}

The general system architecture for this project is highlighted in Figure \ref{fig:arch}. However, specifics about the reasoning behind use and output for individual components are available below. At a high level, the methodology for this research project underwent substantial revisions to enhance data preprocessing and model training efficiency. Initially, the signal separation was performed using the Mad-TwinNet system \cite{twinnet}, followed by Mel-frequency cepstral coefficient (MFCC) feature extraction and classification using a Time-Delay Neural Network (TDNN) and Long Short-Term Memory (LSTM) network. However, to improve scalability and streamline the preprocessing pipeline, the revised final methodology below incorporates a more robust and efficient approach.

\subsection{Signal Separation}
\label{ssec:signal}

The datasets used include a combination of concert recordings and studio recordings. All of these recordings contain a vocalist as the main artist, but some also have background instruments, such as a Mridangam, Violin, or Tambura, in the background. After significant consideration, this background noise was included in the program, to train the program to anticipate the possibility of background noise. Instead, the first ten percent and last ten percent of each clip were trimmed to cut out any spoken introduction to the piece that could add artifacts to the training set.

\subsection{Preprocessing / Feature Extraction}
\label{ssec:mfcc}

The preprocessing pipeline for the Carnatic music dataset is designed to prepare the audio data for feature extraction and model training. One key aspect of this pipeline involves customizing the feature extraction process to capture the nuanced melodic structures and tonal intricacies characteristic of Carnatic music. Rather than relying on standard Mel-frequency cepstral coefficients (MFCCs), the methodology adopts a unique approach, crafting custom bins with triangular filter banks. Each bin represents one of the classical Western notes, spanning a total of 56 bins from B1 to E6. This represents a range inclusive of the standard reach of both male and female voices \cite{vocalrange}. This customization aims to align the feature representation with the distinctive tonal palette of Carnatic music and maximize the capture of the nonstandard melodic patterns inherent in Carnatic ragas.

In addition to customizing the MFCC computation, the preprocessing pipeline incorporates meticulous filtering of frequency components to focus on the essential spectral characteristics relevant to Carnatic music. By significantly attenuating frequencies below or above the specified bins corresponding to classical Western notes, the pipeline ensures that extra extracted artifacts are not included in the feature set. The preprocessing pipeline also standardizes the audio data format and ensures uniformity in segment duration to facilitate consistent processing and analysis. Each audio segment, extracted from the original recordings, undergoes conversion to the WAV format and adjustment of the sample rate to 22050 Hz. Furthermore, the duration of each segment is standardized to the target duration of 30 seconds, with silent padding applied when necessary. These preprocessing steps allow for robust feature extraction and model training, enabling machine learning algorithms to leverage the enriched dataset to accurately classify ragas.

\subsection{Time-Delay Neural Network}
\label{ssec:tdnn}

A Time-Delay Neural Network (TDNN) is utilized as the initial neural network in order to tailor the classifier on the Carnatic music samples. A TDNN was chosen due to its adeptness at capturing temporal dependencies within sequential data, making them well-suited for processing time-series data such as audio signals. Given that Carnatic music — with its near continuous flow of gamakas — exhibits several intricate temporal and microtonal patterns that could make it easily confused with a parallel raga in another shruti, the convolutional layers will be utilized to extract local and translation-invariant features from the spectrogram. These layers will capture patterns in the frequency domain that are indicative of the raga regardless of pitch. Further, TDNNs offer a streamlined architecture that facilitates efficient learning from sequential data while maintaining computational tractability. This characteristic is particularly advantageous for our task, as it allows for the exploration of complex temporal relationships within the music samples without excessive computational overhead. Levering the TDNNs here will allow for the establishment of a strong baseline for comparison with more complex models, such as the Long Short-Term Memory networks used later. As such, there are an $n$ number of convolutional layers, along with additional features that enhance the performance and stability of the neural network models.

These additional layers include batch normalization (BaN) and Rectified Linear Units (ReLU), which further enhance the performance and stability of the neural network models. Batch Normalization is employed to address internal covariate shift within the network during training \cite{batchnormalization}. By normalizing the activations of each layer across mini-batches, Batch Normalization reduces internal covariate shift, thereby stabilizing and accelerating the training process \cite{batchnormalization}. The ReLU activation function is then chosen for its ability to mitigate the vanishing gradient problem commonly encountered in deep neural networks \cite{relu}. By allowing only positive values to pass through, ReLU accelerates convergence during training by promoting sparse activation, thereby facilitating faster learning and more efficient model training \cite{sparserelu}.

Finally, max pooling will be applied to reduce the dimensionality of the output function set, and keep only the most salient features \cite{pooling}. 

\subsection{Long-Short Term Memory}
\label{ssec:lstm}

A Long Short-Term Memory (LSTM) network is utilized in the next step of the classification process. These variations of recurrent neural networks (RNNs) are augmented with specific dense layers and an attention mechanism to approach the task of classifying the ragas from the many samples recorded in different shrutis in the original music dataset. An LSTM is selected for its adeptness at effectively modeling sequential dependencies, crucial for capturing the intricate temporal patterns and melodic structures inherent in music. 

In addition, a flattening layer is introduced after the LSTM module. This flattening layer reshapes the output from the LSTM layer into a one-dimensional array, facilitating the transition from the LSTM's sequential output to the subsequent fully connected layers. The flattened representation retains the learned temporal dependencies while preparing the data for further processing by the dense layers.

Furthermore, multiple dense layers with dropout regularization are integrated into the architecture. Dropout layers are inserted after each dense layer to mitigate overfitting by randomly dropping a fraction of the neurons during training \cite{dropoutlayer}. This regularization technique helps prevent the model from relying too heavily on specific features or relationships in the data, thereby improving its generalization performance. The dense layers, combined with dropout regularization, contribute to the extraction of abstract features from the sequential input data, enhancing the network's ability to classify ragas accurately.

Finally, a softmax layer is employed as the final output layer in the neural network architecture for raga identification in Carnatic music samples. This layer transforms the network's raw outputs into probabilities across different raga classes. By providing probabilistic outputs, the softmax layer enables efficient training via well-defined loss functions and multi-class classification tasks, while also enhancing the transparency and reliability of the raga identification system.

\section{Dataset}
\label{sec:data}

In this project, all data is collected manually through Youtube. Samples were chosen through the following method. First, for each of the 172 ragas, the online database Karnatik.com was perused \cite{karnatik}, and between 1 and 3 songs were selected, depending on the popularity of the raga. Next, each song was searched on YouTube, and two ideal recordings were selected for each song, one in a male shruti and the other in a female shruti. For the purposes of this project, and ideal recording meets all of the following:
\begin{itemize}
    \item Good Sound Quality
    \item More than 4 Minutes
    \item Less than 20 Minutes
    \item No \textit{thani avarthanam}, a prolonged percussion solo \cite{thani}
\end{itemize}

The recordings were chosen as the first in each gender to be listed by YouTube that meet these ideal requirements, regardless of the artist or venue.

In the event that no suitable songs were found under the ideal conditions, relaxations were made to aid in finding a suitable song. These relaxations occurred in the numerical order listed below until a suitable song was found.
\begin{enumerate}
    \item Song between 3 and 4 minutes
    \item Song between 20 and 40 minutes
    \item Ideal song by the other gender
    \item Song between 40 and 60 minutes
    \item Song by the other gender between 3 and 60 minutes
    \item Different song by the same gender in the same raga that hasn't already been sampled and meets the ideal requirements
\end{enumerate}

Under these conditions, a total of 676 samples were chosen across 172 ragas. 62 ragas had 6 samples, 42 ragas had 4 samples, and the remaining 68 ragas had 2 samples. These samples, split into 30-second clips, provided a total of 10,999 clips to use for training and testing. This was split, with 80\% of the data used for training, and 20 \% used for testing, in line with existing methodology \cite{gholamy}.

\begin {table*}[t]
\begin{tabular*}{\textwidth}{@{\hskip 6pt\extracolsep{\stretch{1}}}*{4}{r}}
    \toprule
Reference & Dataset & Classifier & Accuracy \\
    \midrule
Pillai \& Mahajan (2008) \cite{pillai} & All 72 melakartas & Acoustic Model & Vocal, 80.56\% \\
Shetty \& Achary (2009) \cite{shetty} & Arohana-avarohana clips from 20 ragas & ANN & 95\% \\
Ranjani et al. (2011) \cite{ranjani} & 5 Melakarta ragas of Carnatic Music & GMM & 62.10\% \\
Koduri et al. (2011) \cite{koduri2011survey} & 176 clips from 10 ragas & K-NN & 76.50\% \\
Manjabhat et al. (2017) (1) \cite{table} & 213 clips of 14 ragas & FFNN & 90.14\% \\
Manjabhat et al. (2017) (2) \cite{table} & 538 clips from 17 ragas & FFNN & 95\% \\
Chinthapenta (2019) \cite{chinth} & 7 ragas, 20 samples of each raga & LSTM & 30\% \\
This Method & 172 ragas, 676 songs & LSTM/TDNN & 95.31\% \\
    \bottomrule
\end{tabular*}
\caption{Comparison to Previous Work}
\label{table:2}
\end{table*}

\section{Experiment}
\label{sec:exp}

\begin{figure}[t]
    \centering
    \includegraphics[width=0.5\textwidth]{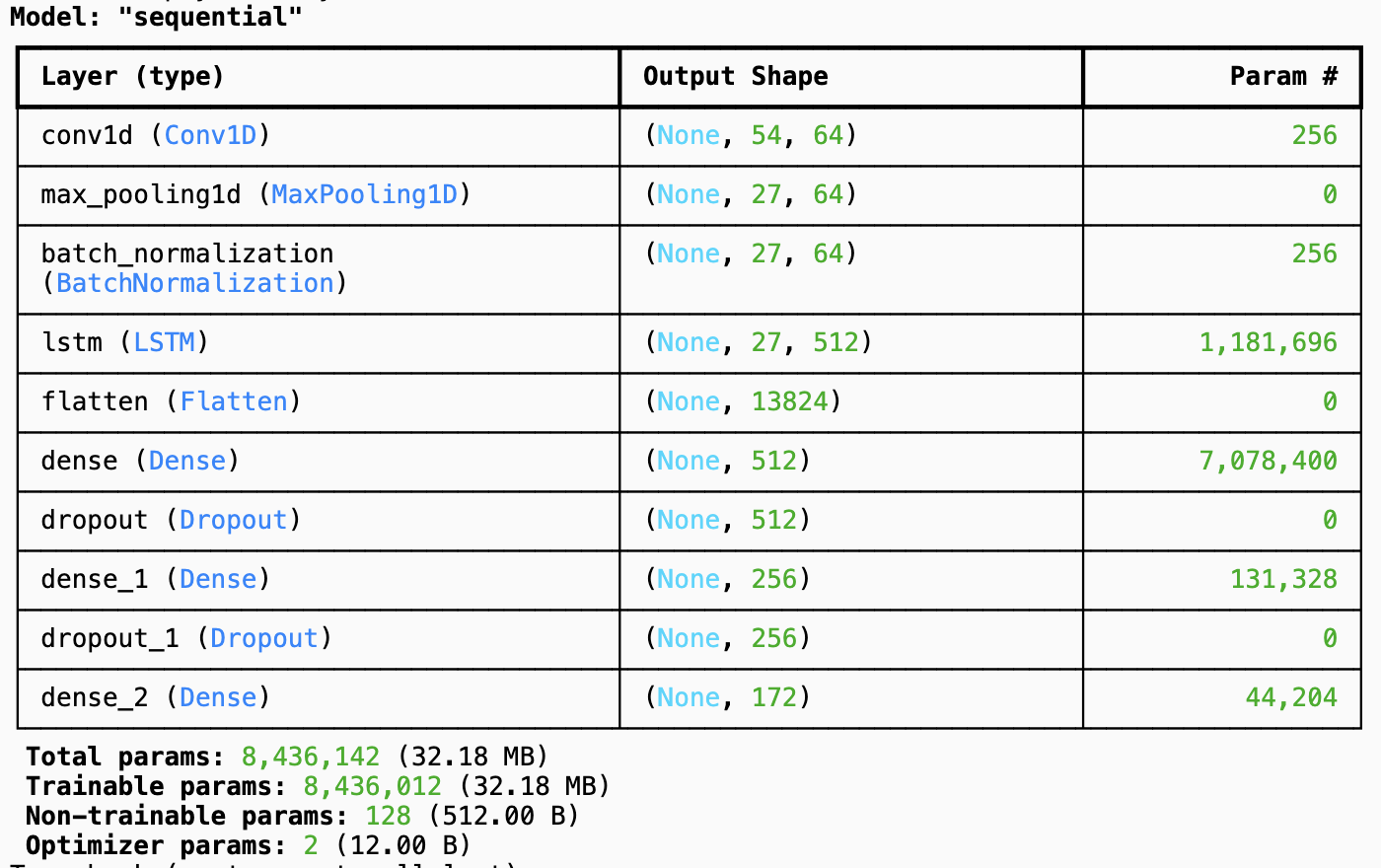}
    \caption{Model Architecture and Weights}
    \label{fig:params}
\end{figure}

The experimental setup for this research paper involves training a deep learning model with a Sequential architecture using Tensorflow, as shown in Figure \ref{fig:params}. The model's architecture comprises of several layers stacked sequentially to form a hierarchical representation of the input data. Specifically, it includes a Convolutional 1D layer with 64 filters and a kernel size of 3, followed by a MaxPooling1D layer to reduce spatial dimensions. BatchNormalization is applied to normalize the activations of the previous layer, enhancing the stability and speed of training. The model further incorporates a Long Short-Term Memory (LSTM) layer with 512 units, which allows it to capture temporal dependencies in the data effectively. Subsequently, a Flatten layer is employed to transform the output into a one-dimensional vector before passing it through two Dense layers with 512 and 256 units, respectively. Dropout layers with a dropout rate of 0.5 are inserted to mitigate overfitting during training. Finally, a Dense layer with 172 units, corresponding to the number of classes in the classification task, produces the model's output probabilities.

The model also has a total of 8,436,142 trainable parameters, spread across the various layers. Additionally, 128 parameters stem from BatchNormalization and two parameters from the optimizer, totaling 8,436,142 parameters. The dataset used for training, validation, and testing is not explicitly described in terms of size or quality. However, it is standard practice to split the dataset into training and test sets, with approximately 80\% allocated for training and 20\% for validation and testing \cite{gholamy}. The quality of the data is influenced by preprocessing techniques, data augmentation strategies, and the representativeness of the dataset with respect to the problem domain.

Training the model involves compiling it with a categorical cross-entropy loss function and the Adam optimizer. To prevent overfitting, early stopping with patience of 100 epochs is employed. The model undergoes training for 300 epochs using a batch size of 256 samples. The backend used for deep learning computations is built using TensorFlow on the Keras layers. This model is specifically designed for a classification task with 172 distinct classes, each representing a unique raga.

\section{Results}
\label{sec:res}

Performance metrics of this model were meticulously assessed, leveraging accuracy and loss computations as benchmarks against previous trials.

\begin{figure}[t]
    \centering
    \includegraphics[width=0.5\textwidth]{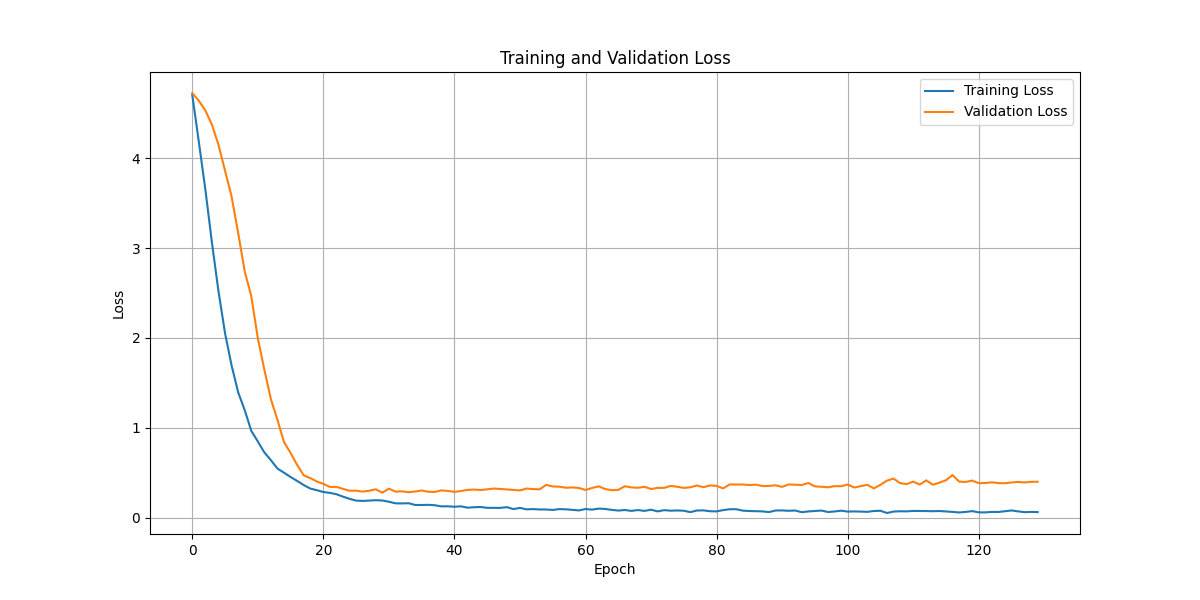}
    \caption{Model Training and Validation Loss}
    \label{fig:loss}
\end{figure}

\begin{figure}[t]
    \centering
    \includegraphics[width=0.5\textwidth]{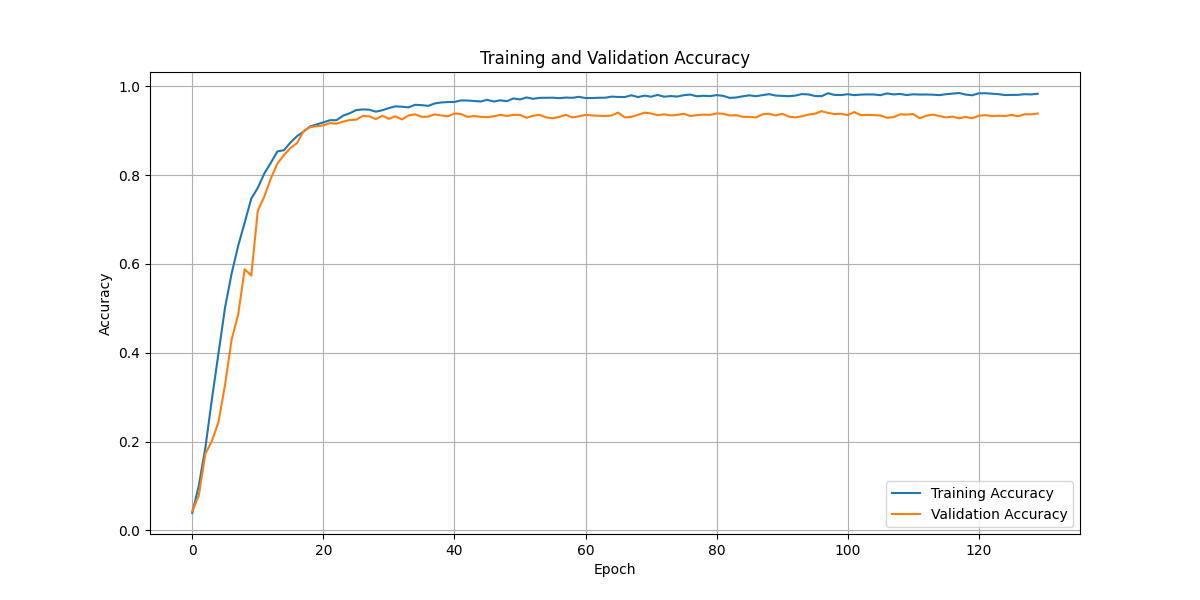}
    \caption{Model Training and Validation Accuracy}
    \label{fig:acc}
\end{figure}

As depicted in Figures \ref{fig:loss} and \ref{fig:acc}, the model's performance on a singular run is shown, revealing some insights. Notably, the implementation of EarlyStopping culminated in the program ending after 132 epochs of the allowed 300. At the time of halt, the recorded final validation loss stood at 0.3544, higher than the training loss of 0.0153, but relatively low given the high data sample and complexity of the task. The EarlyStopping monitor was successful in observing the delta validation loss, stopping exactly one hundred epochs after flat-lining at the $30^{th}$ epoch, clearly in line with its set patience, and preventing an unnecessary continuation of the program through to the $300^{th}$ epoch.

The analysis also delved into the accuracy landscape. The validation accuracy of the model ended at 95.31\%, with a maximum of 96.12\%. Despite trailing marginally behind the training accuracy of 99.57\%, the discrepancy did not suggest any severe risk of overfitting. Although direct comparisons can't be made given the difference in this custom dataset, generalizations can be made. Building off a previous Table published \cite{table}, we can see that our model is on par or better than previous experiments, even with a much more complex dataset with more ragas included, as can be seen in Table \ref{table:2}

\section{Conclusions}
\label{sec:conc}

This study focuses on developing an effective system for identifying Carnatic ragas using advanced machine-learning techniques. By combining Time-Delay Neural Networks (TDNNs) and Long Short-Term Memory (LSTM) networks, we aim to accurately classify ragas based on their distinctive melodic patterns and rhythmic structures.

For this model, a carefully curated dataset of 676 recordings spanning 172 ragas, incorporating both concert and studio performances was utilized. Custom triangular filter banks for feature extraction and attention mechanisms to account for variations in shruti were employed to capture the nuanced characteristics of Carnatic music.

Experimentation with the model yielded promising results, with a validation accuracy of 95.31\%. The implementation of EarlyStopping ensured efficient training and prevented overfitting. The research represents a significant advancement in computational musicology, providing a valuable tool for exploring and preserving the rich heritage of Carnatic music. The results achieve equal or improve on many other past forays in the field, while increasing the number of samples by more than 2 orders of magnitude, and increasing the ragas well beyond the number used in any prior experiments, which are usually measured on a select few popular ragas, or were restricted to the 72 Melakartas. 

Moving forward, we plan to continue refining the methodology and expanding the dataset to encompass a broader range of ragas, shrutis, and styles. Ultimately, the goal is to continue pushing the boundaries of computational music analysis through the lens of Carnatic music and improve accuracy while increasing complexity.



\bibliographystyle{IEEEtran}
\bibliography{ms}

\end{document}